\documentclass[usenatbib]{basi}
\usepackage[T1]{fontenc}
\usepackage[british]{babel}
\usepackage{txfonts}
\usepackage{alltt}
\begin{document}
\title{A colour scheme for the display of astronomical intensity images}

\author[D.~A.~Green]{D.~A.~Green\thanks{email: \texttt{dag@mrao.cam.ac.uk}}\\
                     Cavendish Laboratory,
                     19 J.~J.~Thomson Avenue,
                     Cambridge CB3 0HE, U.K.}

\date{Received --- ; Accepted ---}
\pubyear{2011}
\volume{39}
\pagerange{\pageref{firstpage}--\pageref{lastpage}}

\maketitle
\label{firstpage}

\begin{abstract}
\noindent I describe a colour scheme that is appropriate for the screen display
of intensity images. This -- unlike many currently available schemes -- is
designed to be monotonically increasing in terms of its perceived brightness.
Also, when printed on a black and white postscript printer, the scheme results
in a greyscale with monotonically increasing brightness. This scheme has
recently been incorporated into the radio astronomical analysis packages
\textsc{CASA} and \textsc{AIPS}.
\end{abstract}


\section{The problem}

Images in astronomy often, but not always, represent the intensity of some
source. However, the colour schemes used to display images are not perceived as
increasing monotonically in brightness, which does not aid the interpretation
of the images. The perceived brightness of red, green and blue are not the
same, with green being seen as the brightest, then red, then blue. For example
a bright yellow (i.e.\ full intensity red and green) is perceived as being very
much brighter than a bright blue. So if a colour scheme has yellow for
intermediate intensities, but blue or red for higher intensities, then the blue
or red is perceived at lower brightness. This can be also seen when such colour
images are printed in black and white, when increasing intensity in the image
does not correspond to a greyscale with monotonically increasing brightness.
This problem was noted by \citet{2002IAPM...44...94R} for the colour schemes
then available in the \textsc{Matlab}\footnote{see:
\texttt{http://www.mathworks.com/products/matlab/}} computing package.
\citeauthor{2002IAPM...44...94R} constructed -- in an \emph{ad hoc} manner -- a
colour scheme\footnote{see:
\texttt{http://www.mathworks.com/matlabcentral/fileexchange/2662-cmrmap-m}}
that does result in a greyscale with monotonically increasing brightness when
printed on a black and white device.

Here I describe colour schemes for the display of images that take into account
the different perceptions of the brightnesses of red, green and blue, in order
to maintain a monotonically increasing perception of intensity.

\section{Background}\label{sec:background}

An example of where colour needs to be converted to black and white, to
preserve perceived brightness, is the addition of colour to black and white
television. In the US, the National Television System Committee (NTSC)
specifications from 1953, red, green and blue (hereafter $R$, $G$, $B$) are
mapped to a $Y$ `luma' value (i.e.\ the black and white brightness) signal
using (see, for example, \citealt{Lee:book} or \citealt{Hunt:book})
\begin{equation}
     Y = 0.30 R + 0.59 G + 0.11 B  \label{eq:ntsc}
\end{equation}
(In addition there were two other components, `$Q$' and `$I$' which encoded the
colour.)

The coefficients in equation~\ref{eq:ntsc} are appropriate for the colour
phosphors then in use, and reflect the perceived intensity of the different
colours. Modern monitors differ somewhat, but the coefficients of the different
perceived intensities are similar, e.g.\ the European PAL colour TV standard
uses very similar coefficients, with $Y = 0.299 R + 0.587 G + 0.114 B$, and
more recently HDTV uses $Y=0.2126 R + 0.7152 G + 0.0722 B$ (ITU-R
Recommendation 709, originally from 1993). In all cases the perceived
brightness of green is largest, then red, then blue.

The NTSC coefficients are also used: (1) to map from the \verb|DeviceRGB| to
\verb|DeviceGray| colourspace used within Postscript if the \verb|colorimage|
operator is used with a black and white device\footnote{PostScript Language
Reference, Third Edition, see:
\texttt{http://www.adobe.com/devnet/postscript/}}, and (2) to convert from
colour to greyscale for greyscale only devices, within the \textsc{PGPLOT}
package\footnote{Section~5.1 of the \textsc{PGPLOT} manual, see:
\texttt{http://www.astro.caltech.edu/\string~tjp/pgplot/contents.html}}.

\begin{figure}
\centerline{\includegraphics[width=13cm,clip=]{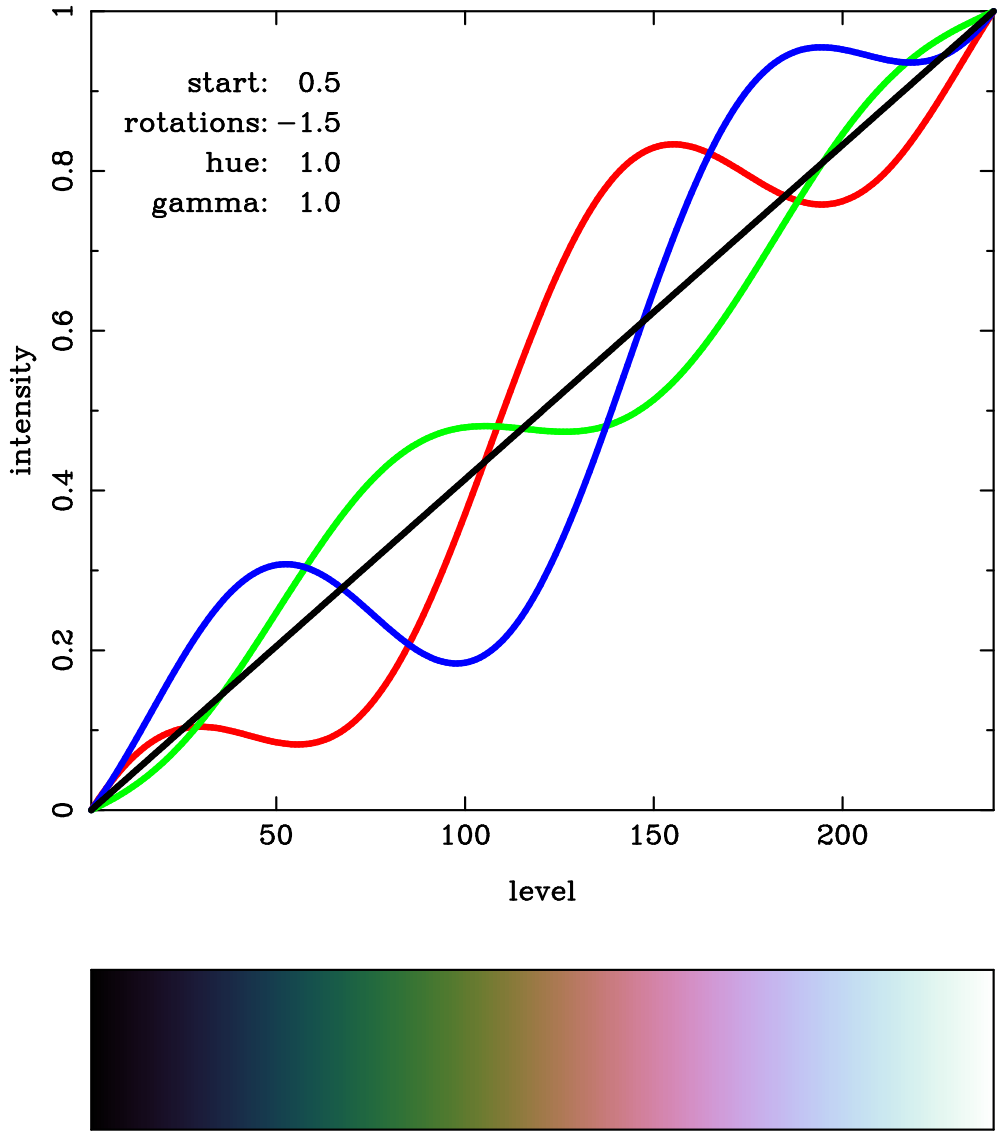}}
\caption{An example `cubehelix' colour scheme with 256 colour levels, as
recently added in \textsc{CASA}. (top) shows the variation in red, green and
blue intensity, and the perceived intensity according to
equation~\ref{eq:ntsc}. (bottom) is a colour wedge to illustrate the scheme.
This has a start colour of 0.5, i.e.\ purple (which is between $R=1$ and
$B=3\equiv0$ when using modulo 3 arithmetic, with $R=1$, $G=2$, $B=3$), with
$-1.5$ rotations, i.e.\ $\to B \to G \to R \to B$, a hue value of 1.2 (as with
the chosen start colour and rotations this value does not lead to any
clipping), and gamma of 1.0.\label{fig:cubehelix}}
\end{figure}

\section{A Solution}\label{sec:solution}

In the 1980s, a colour scheme was implemented for the Sigma ARGS graphics
display on the Cambridge STARLINK VAX, by a colleague, John Fielden. This was
from black to white, spiralling around the greyscale diagonal in an $R$, $G$,
$B$ colour cube. However, this treated $R$, $G$, $B$ equivalently, i.e.\ not
taking into account these colours are perceived differently in terms of their
brightness. I adapted this colour scheme so that it \emph{is} monotonically
increasing in terms of perceived brightness, according to
equation~\ref{eq:ntsc}.

The colour scheme -- a squashed helix around the diagonal of a colour cube,
`cubehelix' -- is implemented as follows. For a unit colour cube (i.e.\ 3-D
coordinates for $R$, $G$, $B$, each in the range 0 to 1), the colour scheme
starts at $(R,G,B)=(0,0,0)$, i.e.\ black, and finishes at $(R,G,B)=(1,1,1)$,
i.e.\ white. For some fraction $\lambda$ between 0 and 1, the colour is the
corresponding grey value at that fraction along the black to white diagonal,
i.e.\ ($\lambda$, $\lambda$, $\lambda$), plus a colour element. The colour
element is calculated in a plane of constant perceived intensity, which can be
defined in terms of two unit $(R,G,B)$ colour vectors $(-0.074, -0.146, 0.986)$
and $(0.891, -0.453, 0.0)$, the first of which is predominantly blue. The
additional colour element of the scheme is controlled by three parameters.
\begin{itemize}
\item The start colour ($s$). This is the direction of the predominant colour
deviation from black at the start of the colour scheme.
\item The number of $R$, $G$, $B$ rotations ($r$) in colour that are made from
the start (i.e.\ black) to the end (i.e.\ white) of the colour scheme.
\item A hue parameter ($h$), which controls how saturated the colours are. The
amplitude of the colour deviation from the grey diagonal is zero at each end
and largest in the middle. If this parameter is zero then the colour scheme is
purely a greyscale, increasing in brightness. If the parameter is too large,
then $R$, $G$, or $B$ may be out of the 0 to 1 range near start or end of
colour scheme, so will have to be clipped (although if only a few colour levels
are clipped, the resulting colour scheme, may still be satisfactory).
\end{itemize}
In addition, a `gamma factor' can be used to provide mapping of $\lambda$ to
$\lambda^\gamma$, to emphasise either low intensity values ($\gamma < 1$), or
high intensity values ($\gamma > 1$). For a fraction $\lambda$ between 0 and 1,
then the $R$, $G$, $B$ values are given by
\begin{equation}
   \pmatrix{ R \cr G \cr B } = \pmatrix{ \lambda^\gamma \cr
                                         \lambda^\gamma \cr
                                         \lambda^\gamma } +
                          a \pmatrix{ -0.14861 & +1.78277 \cr
                                      -0.29227 & -0.90649 \cr
                                      +1.97294 &  0       }
                            \pmatrix{ \cos \phi \cr \sin \phi}
   \label{eqn:RGB}
\end{equation}
where $\phi=2\piup(s/3+r\lambda)$ and $a = h\lambda^\gamma(1-\lambda^\gamma)/2$
(i.e.\ the angle and amplitude for the deviation from the black to white
diagonal) in the plane of constant perceived intensity). Appendix~A gives
\textsc{Fortran 77} code for this algorithm, and Fig.~\ref{fig:cubehelix} shows
one example of a colour scheme constructed using this algorithm.

Note that printed versions of this colour scheme may not be perceived as
monotonically increasing in brightness. This is because the range of colours
that can be represented by colour printers (their `gamut') is usually poorer
than the range that can be represented in computer displays (e.g.\ see
\citealt{Hunt:book}). However, this colour scheme should print as a greyscale
with monotonically decreasing density on black and white postscript devices.

A version of this colour scheme, as shown in Fig.~\ref{fig:cubehelix}, has
recently been incorporated into the \textsc{viewer} task in version 3.1.0 of
the `Common Astronomy Software Application', \textsc{CASA}\footnote{see:
\texttt{http://casa.nrao.edu/}} (e.g.\ \citealt{2008ASPC..394..623J}), called
\texttt{CubeHelix}. This colour scheme has also been recently added to the
31DEC10 version of the `Astronomical Image Processing System',
\textsc{AIPS}\footnote{see: \texttt{http://www.aips.nrao.edu/}} (e.g.\
\citealt{2003ASSL..285..109G}) via the \texttt{TVHELIX} verb, which allows
interactive control of start colour, number of rotations (and direction), the
hue parameter and gamma factor using the \textsc{AIPS} TV buttons and cursor
position.

\section{Discussion and Conclusion}

It should be noted that the colour scheme described in
Section~\ref{sec:solution} is only one example which is suitable for the
display of intensity images. Clearly others could be constructed to take into
account the difference in perception of colours discussed in
Section~\ref{sec:background}. Colour is not only for the display of intensity
images, but is also used in other situations, for example: (i) for polarisation
images, where intensity or polarised intensity is represented by an underlying
greyscale, with polarisation angle with added colour, or (ii) for spectral line
observations, where intensity (or, for example, column density) is represented
by an underlying greyscale, with colour represented radial velocity. In these
cases, ideally the addition of different colours should not distort perception
of intensity also intended. A further subtlety is that the perception of
differences between similar colours is not the same for all colours (e.g.\
\citealt{1943JOSA...33...18M, 1943JOSA...33..675M}), and hence an appropriate
mapping between colour and polarisation angle/velocity is needed.

\section*{Acknowledgements}

I am grateful to John Fielden for his original colour scheme, to several
colleagues for useful discussions, and for Eric Greisen for making this colour
scheme available to \textsc{AIPS} users.


\newpage
\appendix
\section{\textsc{Fortran 77} code}\label{App:code}

This is \textsc{Fortran 77} code that implements the colour scheme as described
in Section~\ref{sec:solution}.

\begin{small}
\begin{alltt}
C=====================================================================72
C Calculates a "cube helix" colour table. The colours are a tapered
C helix around the diagonal of the [R,G,B] colour cube, from black
C [0,0,0] to white [1,1,1] Deviations away from the diagonal vary
C quadratically, increasing from zero at black, to a maximum, then
C decreasing to zero at white, all the time rotating in colour.
C
C The input parameters controlling the colour helix are:
C
C    START colour (1=red, 2=green, 3=blue; e.g. 0.5=purple);
C    ROTS  rotations in colour (typically -1.5 to 1.5, e.g. -1.0
C          is one blue->green->red cycle);
C    HUE   for hue intensity scaling (in the range 0.0 (B+W) to 1.0
C          to be strictly correct, larger values may be OK with
C          particular start/end colours);
C    GAMMA set the gamma correction for intensity.
C
C The routine returns a colour table NLEV elements long in RED, GRN
C and BLU (each element in the range 0.0 to 1.0), and the numbers,
C NLO and NHI, of red, green or blue values that had to be clipped
C because they were too low or too high.
C---------------------------------------------------------------------72
C Dave Green --- MRAO --- 2011 June 13th
C---------------------------------------------------------------------72
C See:
C   Green, D. A., 2011, Bulletin of the Astronomical Society of India,
C      Vol.39, p.\pageref{firstpage}
C---------------------------------------------------------------------72
      SUBROUTINE CUBHLX(START,ROTS,HUE,GAMMA,NLEV,RED,GRN,BLU,NLO,NHI)
C     ================================================================
C
      INTEGER   NLEV,I,NLO,NHI
      REAL      START,ROTS,HUE,GAMMA
      REAL      RED(NLEV),GRN(NLEV),BLU(NLEV)
      REAL      PI,FRACT,ANGLE,AMP
C
      PI=4.0*ATAN(1.0)
      NLO=0
      NHI=0
C
      DO 1000 I=1,NLEV
        FRACT=FLOAT(I-1)/FLOAT(NLEV-1)
        ANGLE=2*PI*(START/3.0+1.0+ROTS*FRACT)
        FRACT=FRACT**GAMMA
        AMP=HUE*FRACT*(1-FRACT)/2.0
        RED(I)=FRACT+AMP*(-0.14861*COS(ANGLE)+1.78277*SIN(ANGLE))
        GRN(I)=FRACT+AMP*(-0.29227*COS(ANGLE)-0.90649*SIN(ANGLE))
        BLU(I)=FRACT+AMP*(+1.97294*COS(ANGLE))
C
        IF(RED(I).LT.0.0)THEN
          RED(I)=0.0
          NLO=NLO+1
        ENDIF
        IF(GRN(I).LT.0.0)THEN
          GRN(I)=0.0
          NLO=NLO+1
        ENDIF
        IF(BLU(I).LT.0.0)THEN
          BLU(I)=0.0
          NLO=NLO+1
        ENDIF
C
        IF(RED(I).GT.1.0)THEN
          RED(I)=1.0
          NHI=NHI+1
        ENDIF
        IF(GRN(I).GT.1.0)THEN
          GRN(I)=1.0
          NHI=NHI+1
        ENDIF
        IF(BLU(I).GT.1.0)THEN
          BLU(I)=1.0
          NHI=NHI+1
        ENDIF
 1000 CONTINUE
C
      RETURN
      END
\end{alltt}
\end{small}

\label{lastpage}
\end{document}